\documentclass{emulateapj}
\usepackage{graphicx}
\usepackage{hyperref}
\usepackage{natbib}
\usepackage{float}
\begin{document}
\nocite{*}
\title{Rich: Open Source Hydrodynamic Simulation on a Moving Voronoi Mesh}
\author{Almog Yalinewich$^1$, Elad Steinberg$^1$ and Re'em Sari$^{1,2}$}
\affiliation{$^1$Racah Institute of Physics, the Hebrew University, 91904,
Jerusalem, Israel \\
$^2$California Institute of Technology, MC 130-33, Pasadena, CA
91125}

\begin{abstract}
\noindent
We present here RICH, a state of the art 2D hydrodynamic
code based on Godunov's method, on an unstructured moving mesh (the acronym stands for Racah Institute Computational Hydrodynamics). This code is largely based on the code AREPO. It differs from AREPO in the interpolation and time advancement scheme as well as a novel parallelization scheme based on Voronoi tessellation. Using our code we study the pros and cons of a moving mesh (in comparison to a static mesh). We also compare its accuracy to other codes.
Specifically, we show that our implementation of external sources and time advancement scheme is more accurate and robust than AREPO's, when the mesh is allowed to move. We performed a parameter study of the cell rounding mechanism (Llyod iterations) and it effects. We find that in most cases a moving mesh gives better results than a static mesh, but it is not universally true. In the case where matter moves in one way, and a sound wave is traveling in the other way (such that relative to the grid the wave is not moving) a static mesh gives better results than a moving mesh. Moreover, we show that Voronoi based moving mesh schemes suffer from an error, that is resolution independent, due to inconsistencies between the flux calculation and change in the area of a cell. Our code is publicly available as open source and designed in an object oriented, user friendly way that facilitates incorporation of new algorithms and physical processes. 
\end{abstract}

\section{Introduction}

It has long been recognized that the aid of computers can greatly
increase our understanding of astrophysical phenomena. And yet even with the
progress of computers, solutions to some problems are still limited by 
computing power. One idea to achieve greater accuracy at
a given computer power, is using a computational mesh that moves together
with the fluid (Lagrangian grid), rather than the more common static
mesh (Eulerian grid). While it has not been proven that the former
is better, one reason for using a Lagrangian grid
is that it automatically gets denser (thus providing higher resolution)
in places where matter is flowing into (e.g. behind shock fronts).
Since these areas are usually the more interesting parts of the domain,
the Lagrangian grids tend to give better resolution in areas of interest.

Recently, a novel method for a semi-Lagrangian Gudonov scheme,
called AREPO \citep{Springel2010}, was published. In AREPO, in contrast to ALE (Arbitrary Lagrangian Eulerian)
simulations, mesh points are no longer required to adhere to their
neighbors, so when computational cells drift too far apart, they do
not tangle but change neighbors. Another advantage of this scheme
is that each flux is calculated in the reference frame of the moving
edge, so advections between cells are greatly reduced. AREPO is thus
able to reap most of the benefits of ALE without any of the drawbacks.

Despite a comprehensive description of the code and its
test suite in \cite{Springel2010}, a more thorough comparison between
semi-Lagrangian and Eulerian grids is in order. Another matter that
requires more work is the coupling of external forces/sources. While
the method employed in \cite{Springel2010} to couple gravity is explained
in detail, there is no simple way to extend it to arbitrary sources terms.  We developed our own version of the code, called RICH, which
is written in C++ and takes after AREPO and its relativistic variant
TESS \citep{Duffell2011}, with a few changes.
The purpose of this paper is to present our code, compare its accuracy with other codes, do a parameter study of the mesh rounding mechanism (Llyod iterations) and to explore the pros and cons of using a semi-Lagrangian grid. Since most of the algorithms were discussed in other papers
\citep{Springel2010,Duffell2011}, in this paper we will focus on
the differences between our code and AREPO and TESS.

  The paper is structured as follows. In
section \ref{sec:difference_from_arepo} we describe the differences
of our code from AREPO and TESS. One and two dimensional test problems are presented
in sections \ref{sec:One-Dimensional} and \ref{sec:Two-dimensional}
respectively. The effect of non-Lagrangian motion is discussed in section \ref{sec:nonlagrangian}. We show that there is a resolution independent error that arises from an inconsistency between the flux calculation and change in the area of a cell in section \ref{sec:error}. The question of whether a Lagrangian code is always better than an Eulerian code is addressed in section \ref{sec:lagbetter}. In section \ref{sec:Conclusion} we summarize and discuss the advantages of using a semi-Lagrangian grid.

\section{Algorithm Modifications}\label{sec:difference_from_arepo}

Since our code is very similar to AREPO \citep{Springel2010} and TESS \citep{Duffell2011} we will only dwell on the differences from them.

\subsection{Tessellation Creation}
\noindent
We follow AREPO and TESS and construct the Voronoi diagram by
first building the Delaunay triangulation (its dual graph) and
then we translate the triangulation to the Voronoi diagram
in linear time. We create the Delaunay triangulation using the 
point insertion method \citep{voronoi,Springel2010}. This method 
adds the mesh generating points one after another, and checks each 
time whether it falls inside a circumscribing circle of an existing
triangle. One difficulty with this stage occurs when
a point lies exactly on the circle, since numerical round off errors 
can change the result. We adopt the method proposed by \cite{Shewchuk1996} 
to use adaptive floating point arithmetic whenever the round-off error in the calculation may change
the sign of the answer. This method was tested on a set of points
arranged in a square grid (so that all of the triangles are degenerate) and
the in-circle test time was about twice that of normal arithmetic.
When the triangulation was tested with a random set of points only
a $10\%$ increase in the triangulation time was observed. 
Constructing a Voronoi diagram with
$10^{6}$ random points takes $6.7$ seconds on an i7-2620M CPU and
a square mesh with the same number of points takes $8.7$ seconds.
For comparison, using the same CPU but using the qhull algorithm with
MATLAB 2013a took 14 seconds, while AREPO reported 516 seconds for building
5 billion mesh points with 1024 SGI Altix 4700 cpus (but in three
dimensions).

\subsection{Interpolation}

Higher order schemes require spatial interpolation of the cell values.
AREPO reconstructs the gradient in each cell using the Green-Gauss theorem and we implement
the same method. Specifically
\begin{equation}
\label{eq:slope}
<\vec{\nabla}\phi>_{i}=\frac{1}{A_{i}}\sum_{i\neq j}L_{ij}([\phi_{j}-\phi_{i}]\frac{\vec{c_{ij}}}{r_{ij}}-\frac{\phi_{i}+\phi_{j}}{2}\frac{\vec{r_{ij}}}{r_{ij}})
\end{equation}
where $\phi_{i}$ is the quantity to reconstruct in the $i$ cell,
$A_{i}$ is the cell's area, $L_{ij}$ is the length of the edge between
the $i$ and $j$ cells, $r_{ij}$ is the vector connecting the two
mesh generating points and $c_{ij}$ is the vector from the midpoint
between the $i$ and $j$ mesh generating points and the center of
the edge between the cells. The summation is done among all of the
cell's neighbors. Once the gradient is known, the primitive variables
at the edges are reconstructed using linear extrapolation
\begin{equation}
\phi_{ij}=\phi_{i}+<\vec{\nabla}\phi>_{i}\cdot\left(\vec{L}_{mid}-\vec{s}_{i}\right)
\end{equation}
where $\vec{s}$ is the cell's center of mass and $\vec{L}_{mid}$ is the middle of the edge.

\begin{figure}
\begin{centering}
\includegraphics[width=0.4\textwidth,,keepaspectratio]{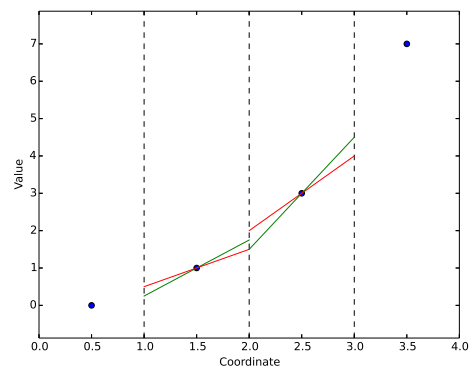}
\par\end{centering}
\caption{Example for TVD violation in AREPO's interpolation. Blue points represent
cell values, dashed black line cell boundaries, green lines the
interpolations according to AREPO, and red lines the TVD interpolation. One can see that AREPO's method creates new local exterma, where as the TVD method does not. \label{fig:tvd_violation}}
\end{figure}

  In order to prevent the creation of new maxima or minima,
which can cause oscillations near discontinuities, a slope limiter
is used. AREPO's slope limiter prevents the creation of a global maxima/minima
in the sense that the extrapolated value cannot exceed the value of
the highest neighbor and cannot be below the lowest neighbor. In order
to achieve this, the gradient is set to be
\begin{equation}
<\vec{\nabla}\phi>_{i}'=\alpha_{i}<\vec{\nabla}\phi>_{i}
\end{equation}
where the slope limiter $\alpha_{i}$ is set to be
\begin{equation}
\alpha_{i}=min(1,\psi_{ij})
\end{equation}
\begin{equation}
\psi_{ij}=\left\{ \begin{array}{c}
\left(\phi_{i}^{max}-\phi_{i}\right)/\Delta\phi_{ij}\\
\left(\phi_{i}^{min}-\phi_{i}\right)/\Delta\phi_{ij}\\
1
\end{array}\quad\begin{array}{c}
\Delta\phi_{ij}>0\\
\Delta\phi_{ij}<0\\
\Delta\phi_{ij}=0
\end{array}\right.
\end{equation}
where $\Delta\phi_{ij}$ is the difference between the interpolated value at the edge and the value at the cell center, and $\phi_{i}^{max}$ and $\phi_{i}^{min}$ are the maximum and minimum
values among the neighbors respectively. This slope limiting is not
Total Variation Diminishing (TVD) \citep{Toro1999}, since under these
constraints, the gradient of interpolated values can have a different
sign from the gradient of the two neighboring mesh points. To demonstrate
this problem, let us assume a uniform 1D grid with 4 equal sized computation
cells, whose values are 0, 1, 3 and 7. Applying the method described
above, we get that although the value of cell \#3 is greater than\emph{
}cell \#2, the interpolated value of cell \#2 is greater than that
of cell \#3, as can be seen in figure \ref{fig:tvd_violation}.

  A possible remedy for this problem is presented in TESS
\citep{Duffell2011}, where they employ a more restrictive ``local''
slope limiter, in which the extrapolated value cannot exceed any of
its neighbors, by some numerical factor $\theta$
\begin{equation}
\psi_{ij}=\left\{ \begin{array}{c}
\max\left(\theta\left(\phi_{j}-\phi_{i}\right)/\Delta\phi_{ij},0\right)\\
1
\end{array}\quad\begin{array}{c}
\Delta\phi_{ij}\neq0\\
\Delta\phi_{ij}=0
\end{array}\right.
\end{equation}

  Choosing $\theta\le0.5$ prevents TVD violations. The tradeoff
is that the lower $\theta$ is, the more diffusive the scheme becomes.

  By default we use AREPO's scheme, unless we suspect that
there's a shock front in the neighborhood of cells, in which case
we use TESS' scheme instead. The quantitative criterion is when either of these two conditions is true
\begin{equation}
-\nabla\mathbf{v}>\delta_{v}\frac{c_{s_{i}}}{R_{i}}
\end{equation}
\begin{equation}
\delta p>\min\left\{ \begin{array}{c}
P_{i}/P_{j}\\
P_{j}/P_{i}
\end{array}\quad\begin{array}{c}
P_{j}>P_{i}\\
P_{i}>P_{j}
\end{array}\right.
\end{equation}
The default numerical factors we chose are $\theta=0.5$, $\delta_{v}=0.2$
and $\delta_{p}=0.7$.

\subsection{Time Advancement}

In order to achieve second order accuracy, AREPO uses linear interpolation to determine the hydrodynamic variables near the edge of a cell at the beginning of a time step. By substituting the spatial derivatives into the hydroydnamic equations it gets the time derivative of the primitive variables on the edge between cells, and uses them to estimate the values at the edge at half a time step. We call this method of time advancement, ``extrapolated fluxes''. Then, the ``time centered'' primitive variables are given to the Riemann solver in order to compute second order accurate fluxes. Adding these fluxes to the conserved variables from the beginning of the time step yield second order accuracy, and only invokes the Riemann solver and interpolations once. The down side is that external sources have to be written in a
special way so they will be second order accurate in time as well.
For instance, gravity has to use the variables from before the time
step and after the time step in order to be second order accurate.

  We use a different approach for our time integration. Following
TESS, we use a ``time centered  fluxes'' scheme. The system is advanced by half a time step (using linear spatial interpolation), the mesh
is rebuilt and the half time step primitive variables are computed. Then, the time centered fluxes are computed and are added to the conserved variables from the beginning of the time step with a full time step. The final mesh is the built from advancing the mesh points a full time step, from their position at the beginning of the time step with a velocity that was calculated during the half time step. This ``time centered  fluxes'' scheme ensures that our time advancement is second order accurate.

The ``time centered  fluxes'' scheme has the added
benefit that external sources have to be only first order accurate
in time and the time integration will automatically make them second
order accurate.
The Courant stability condition also allows us to
use a Courant number larger than unity, though we usually use the
default arbitrary value 0.3. The reason it is so low is to prevent
the simulation from crashing due to other reasons, like the depletion
of energy from a cell due to a strong rarefaction wave.

The down side to our scheme is that it requires
twice the computation time since it requires building the mesh and
calculating the fluxes twice. In many applications the robustness
of the external forces implementation is worth the slower execution
time.

\subsection{HLLC Riemann Solver}

The Riemann solver used in AREPO is exact. This means that calculating
the flux on every edge involves a numerical solution of a single
variable equation \citep{Toro1999}.  The downside of using this solver is
that it is time consuming and that it is generally 
not applicable to all equation of states. To avoid these difficulties, we implemented
the HLLC Riemann solver \citep{Toro1999}. This is an approximate solver,
so it does not return the correct flux when there is a large difference between
between the values of the hydrodynamic variables in adjacent cells. However,
since the Godunov method tends to smear a discontinuity across a few cells (and thus reduce the difference between adjacent cells) the values of the hydrodynamic variables before and after the discontinuity converge to the correct values
within a few time steps. Also, since
it only uses the energy and speed of sound, it can be used with any
equation of state, not just ideal gas.

\subsection{Parallelization Scheme}
Our code is made parallel by the use of the MPI interface. Our domain is decomposed by building a Voronoi diagram from CPU points that represent the different CPUs, and each CPU holds in its memory only the hydro points that are inside its Voronoi cell. In order to maintain a good load balance throughout the run, we move the CPU mesh points in a way to preserve the workload roughly equal. Our parallelization scheme is discussed in Steinberg et al 2014.
\section{One Dimensional Test Problems}\label{sec:One-Dimensional}
In order to test our code, we run a set of 1D and 2D test problems. For all of the 1D test problems we compute the convergence rate and an error function, that is problem specific, by repeating the tests with resolutions of $32-256$ cells. A summary of all our results is given in table \ref{table:convergence}.

\subsection{Simple Waves}

We repeat the test described in \cite{Colella2006}, which tests the
propagation of large perturbations. The density is given by
\begin{equation}
\rho_{0}\left(x\right)=\left\{ \begin{array}{c}
1\\
1+10\cdot\left(\left(x-0.3\right)^{2}-0.5\right)^{4}
\end{array}\quad\begin{array}{c}
\left|x-0.3\right|>0.5\\
\left|x-0.3\right|<0.5
\end{array}\right.
\end{equation}
and the pressure and velocity are chosen so that the entropy and the
negative Riemann invariant would be uniform throughout the domain,
so there would only be a forward moving wave. The pressure and the velocity are given by
\begin{equation}
p_{0}\left(x\right)=\rho_{0}^{\gamma}\left(x\right),
\end{equation}
\begin{equation}
v_{0}\left(x\right)=\frac{2}{\gamma-1}\left[\sqrt{\gamma\frac{p_{0}\left(x\right)}{\rho_{0}\left(x\right)}}-\sqrt{\gamma\frac{p_{0}\left(1\right)}{\rho_{0}\left(1\right)}}\right],
\end{equation}
where $\gamma=\frac{5}{3}$ and the problem is set up with a domain of $\left[0,1\right]$ using periodic boundaries. The calculation was terminated at time $t=0.02$. In this test $L$ is defined as
\begin{equation}
L=\sum_{i=1}^N\frac{\left|\phi^n_i-\phi^a_i\right|}{|\phi^a_i|N}
\label{eq:Lsimple}\end{equation}
where $\phi^n_i$ is the numerical hydrodynamical quantity, $\phi^a_i$ is the analytical value, $N$ is the total number of cells and subscript $i$ represents the value at spatial position with index $i$. For the velocity we replace $\phi^a_i(x)$ in the denominator by the speed of sound. Repeating this comparison at different resolutions, we obtain convergence curves that show that the code is indeed second order accurate; $L\propto N^{-2}$. All schemes achieve a second order convergence rate and we calculate the error prefactor, $A$, as $L=AN^{-2}$.
For this test a semi-Lagrangian grid gives better results than an Eulerian grid by a factor of about 3.5, since the velocities are not small. The extrapolated fluxes time advancement scheme is only slightly ($1\%-8\%$) better than ours.

\subsection{Acoustic Waves}

The acoustic waves problem checks how small perturbations propagate.
When the perturbations are small, the hydrodynamic equations can be
linearized and solved analytically \citep{Landau1987}. Another feature
of small perturbations is that the approximate Riemann solver gives
results wich are very close to the exact Riemann solver \citep{Toro1999}.

The problem is set up with a domain of $\left[0,1\right]$ and equation of state is of an ideal gas
with adiabatic index $\gamma=\frac{5}{3}$. The boundary conditions
are set to be periodic, and the initial conditions are
\begin{equation}
\rho_{0}\left(x\right)=1+10^{-6}\sin\left(2\pi x\right)
\end{equation}
\begin{equation}
p_{0}\left(x\right)=\frac{3}{5}+10^{-6}\sin\left(2\pi x\right)
\end{equation}
\begin{equation}
v_{0}\left(x\right)=0
\end{equation}

We compare the spatial profile at time $t=1$ (the
time it takes a sound wave to go full circle), to the analytical profiles (which are identical to the initial conditions), using
\begin{equation} \label{eqn:l1_definition_no_1}
L=\sum_{i=1}^N\frac{\left|\phi^n_i-\phi^a_i\right|}{10^{-6}N}
\label{eq:Lacoustic}\end{equation}
Since the velocities are very low, there is almost no difference between the Eulerian and Lagrangian schemes, both achieve second order convergence rate and we report only the Eulerian schemes.

The measured prefactors that are reported in table \ref{table:convergence} show that in this test, the extrapolated fluxes time advancement scheme has a prefactor that is about 1.5 lower than the time centered fluxes time advancement scheme.

\subsection{Shock Tube}

The shock tube problem tests the code's ability to resolve strong
shocks and discontinuities. The initial conditions are
\begin{equation}
\rho_{0}\left(x\right)=1,\quad p_{0}\left(x\right)=\left\{ \begin{array}{c}
1\\
10
\end{array}\quad\begin{array}{c}
x>0.5\\
x<0.5
\end{array}\right.,\quad v_{0}\left(x\right)=0
\end{equation}
in the domain $x\in\left[0,1\right]$. The exact, self similar solution can be found by the solution of the Riemann problem \citep{Toro1999} and the profiles are compared to the analytical solution at time $t=0.1$. Since the test involves a strong shock wave, a few cells dominate the error if the error $L$ is defined by equation ~\ref{eqn:l1_definition_no_1},~ and information about the rest of the cells is essentially disregarded. In order to prevent this, we define $L$ as
\begin{equation}
L=\sum_{i=1}^N\frac{|\phi^n_i-\phi^a_i|^{1/4}}{|\phi^a_i|^{1/4}N}.
\label{eq:Lshock}\end{equation}
except for the velocity which is again normalized by the speed of sound.
Second order convergence with the above $L$ translates to $L\propto N^{-1/2}$. Our runs show that our code is indeed second order accurate and the calculated prefactors, given in table \ref{table:convergence} show that the time centered fluxes time advancement scheme has the same error as the extrapolated fluxes scheme and in both schemes the semi-Lagrangian movement is better than Eulerian.
\subsection{Standing Driven Waves}
The main goal of this problem is to test the accuracy of the code
when coupled to external forces/sources. We start out with a smooth
uniform hydrodynamic profile $\rho\left(x\right)=\rho_{0}=1$, $p\left(x\right)=p_{0}=1$,
$v=v_{0}=0$ on the domain $x\in\left[0,1\right]$ with periodic boundary conditions. Perturbations
are introduced by an external acceleration
\begin{equation}
f\left(x,t\right)=A\sin\left(kx\right)\sin\left(ktv\right)
\end{equation}
where $A=10^{-4}$ has the units of acceleration, $k=2\pi$ and $v=0.1$.
The perturbations in the hydrodynamic
variables, which are obtained from the analytical solution for $A \ll 1$, are given by
\begin{equation}
\delta\rho=\frac{A\rho_{0}\cos\left(kx\right)\left(v\sin\left(c_{0}kt\right)-c_{0}\sin\left(ktv\right)\right)}{c_{0}k\left(c_{0}^{2}-v^{2}\right)}
\end{equation}
\begin{equation}
\delta p=\frac{Ac_{0}\rho_{0}\cos\left(kx\right)\left(v\sin\left(c_{0}kt\right)-c_{0}\sin\left(ktv\right)\right)}{k\left(c_{0}^{2}-v^{2}\right)}
\end{equation}
\begin{equation}
\delta v=\frac{Av\left(\cos\left(ktv\right)-\cos\left(ktc_{0}\right)\right)\sin\left(kx\right)}{k\left(c_{0}^{2}-v^{2}\right)}
\end{equation}
where $c_{0}=\sqrt{\gamma p_{0}/\rho_{0}}$ and $\gamma=\frac{5}{3}$.
For this test problem we define the error function as
\begin{equation}
L=\sum_{i=1}^N\frac{|\phi_i^n-\phi_i^a|}{|\phi_i^a|AN}
\label{eq:Ldriven}\end{equation}
and the velocity is once again normalized by the speed of sound.
The measured convergence rates for all of the schemes is second order, and since the velocities are small we report only the Eulerian results. The prefactor for error in density and pressure with the time centered flux time advancement scheme is smaller than the extrapolated fluxes scheme (by about $25\%$) while for the velocity, extrapolated fluxes time advancement is better by a factor of 2.

\section{Two dimensional Test Problems}\label{sec:Two-dimensional}

\subsection{Pure Advection}

One of the benefits of having a moving mesh is that it should handle
advection much better than Eulerian codes. In this test we set the
velocity and pressure to constant values and the density to some non
trivial distribution. Physically, it is equivalent to a static environment
viewed from a moving reference frame. When those initial conditions
are advanced by an Eulerian scheme, the features of the initial density
distribution tend to diffuse. In a Lagrangian scheme we would expect
no such distortion, so the error should be 0, up to numerical precision.
We note that the motion of the mesh generating points slightly deviates
from Lagrangian motion in order to make the cells round. Therefore,
if the initial cells will not be round enough, there will be some
diffusion even in the ``Lagrangian'' case.

We choose the pressure to be $p=1$, velocity $\vec{v}=\hat{x}+\hat{y}$
(so it won't be parallel to either axis) and the density distribution
to be
\[
\rho\left(r\right)=\left\{ \begin{array}{c}
100\\
1
\end{array}\quad\begin{array}{c}
r<0.2\\
r>0.2
\end{array}\right.
\]
The domain is set to be$\left[-0.5,0.5\right]^{2}$ with periodic
boundary conditions. The simulation is then run to time 1 (the time
it takes all the points to come full circle) and compare the initial
and final snapshots of the density. The test was run with different
resolutions and in all cases the errors were consistent with machine
round off error when the mesh was allowed to move with the fluid.

When considering pure advection, the moving mesh has the great advantage
of having zero error, compared to Eulerian codes where the error depends
on the fluid's velocity.

\subsection{Noh Problem}

The Noh problem \citep{NOH1987} checks how the code handles strong
shocks and highly supersonic flow. The setup for the test is a uniform
density $\rho_{0}=1$, small uniform pressure $p=10^{-6}$ and uniform
radial inflow velocity $v=1$ while the adiabatic index is set to $\gamma=\frac{5}{3}$.
The analytic self similar solution is
\begin{equation}
\rho\left(r,t\right)=\left\{ \begin{array}{c}
\left(\frac{\gamma+1}{\gamma-1}\right)^2 \\
1+t/r
\end{array}\quad\begin{array}{c}
r<t\left(\gamma-1\right)/2\\
r>t\left(\gamma-1\right)/2
\end{array}\right.
\end{equation}
\begin{equation}
v\left(r,t\right)=\left\{ \begin{array}{c}
0\\
-1
\end{array}\quad\begin{array}{c}
r<t\left(\gamma-1\right)/2\\
r>t\left(\gamma-1\right)/2
\end{array}\right.
\end{equation}

\begin{equation}
p\left(r,t\right)=\left\{ \begin{array}{c}
\frac{1}{2}\frac{\left(\gamma+1\right)^{2}}{\gamma-1}\\
10^{-6}
\end{array}\quad\begin{array}{c}
r<t/3\\
r>t/3
\end{array}\right.
\end{equation}

 One way to simulate this problem is to have the computational
grid only in the first quadrant $\left[0,1\right]^{2}$ and use rigid
wall boundary conditions on the lower $\left(y=0\right)$ and left
$\left(x=0\right)$ boundaries. However, we choose to take after AREPO
and use the computational domain of $\left[-1,1\right]^{2}$. This
allows us to verify how well our code preserves reflection symmetry, which was achieved.

We use 2500 mesh generating points, randomly distributed
across the domain and the boundary conditions are dictated from the
analytic solution. An inflow boundary condition poses no difficulty
in case of Eulerian point motion, but in case of a semi-Lagrangian point
motion, cells close to the origin would tend to compress and shrink,
thus causing the time step to plummet, while cells far away from the
origin would tend to bloat, thus causing loss of precision. To remedy
this problem we use adaptive mesh refinement, like in AREPO. We split
cells once their volume increases above 150\% of their initial value
and coarsen them when their volume drops below 25\% of their initial
value.

Due to the strong shock wave in this test, we use the same error function as described in the Shock Tube test.
\begin{equation}
L=\sum_{i=1}^N\frac{|\phi_i^n-\phi_i^a|^{1/4}}{|\phi_i^a|^{1/4}N}.
\label{eq:Lnoh}\end{equation}
The results in table \ref{table:convergence} show that all of the schemes give comparable results with Eulerian being slightly better. This is the result of the AMR scheme that de-refines the cells around the shock front and the area inconsistency error that is described in section \ref{sec:error}.

\subsection{Gresho Vortex}
\label{sec:gresho}
We repeated the simulation of the Gresho vortex problem as described in the AREPO code paper \citep{Springel2010}. In this problem, the initial
density is uniform and equal to 1. The pressure is given, in polar
coordinates, by
\begin{equation}
P\left(r,\phi\right)=\left\{ \begin{array}{c}
5+\frac{25}{2}r^{2}\\
9+\frac{25}{2}r^{2}-20r+4\ln\left(5r\right)\\
3+4\ln2
\end{array}\quad\begin{array}{c}
r<\frac{1}{5}\\
\frac{2}{5}>r>\frac{1}{5}\\
r>\frac{2}{5}
\end{array}\right.
\end{equation}
and the azimuthal velocity is
\begin{equation}
v_{\phi}\left(r,\phi\right)=\left\{ \begin{array}{c}
5r\\
2-5r\\
0
\end{array}\quad\begin{array}{c}
r<\frac{1}{5}\\
\frac{2}{5}>r>\frac{1}{5}\\
r>\frac{2}{5}.
\end{array}\right.
\end{equation}
The pressure balances the centrifugal force, so the variables should
not change in time.
In this test $L$ is defined as
\begin{equation}
L=\sum_{i=1}^N\frac{|\phi_i^n-\phi_i^a|}{|\phi_i^a|N}
\label{eq:Lgresho}\end{equation}
Like \cite{Springel2010}, we found $L \propto N^{-1.5}$ and the prefactor in the time centered fluxes time advancement scheme was slightly better than the one in the extrapolated fluxes scheme. With the time centered fluxes time advancement scheme, there was no difference between Lagrangian and Eulerian grid motion.

\begin{table}[p]
\begin{tabular}{|c|c|c|c|c|c|c|}
\hline
Test Name & $\alpha$ & L eq. & \multicolumn{2}{|l|}{Time Centered Fluxes - RICH}&\multicolumn{2}{|l|}{Extrapolated Fluxes - AREPO} \\
\hline
 & & & Eulerian & Semi-Lagrangian & Eulerian & Semi-Lagrangian\\
\hline
Acoustic waves density & -2 & \ref{eq:Lacoustic} & $1.07\cdot10^{-4}$ & - & $7.06\cdot10^{-5}$ & - \\
\hline
Acoustic waves pressure & -2 & \ref{eq:Lacoustic} & $1.07\cdot10^{-4}$ & - & $7.06\cdot10^{-5}$ & - \\ \hline
Acoustic waves velocity & -2 & \ref{eq:Lacoustic} & $1.07\cdot10^{-4}$ & - & $7.06\cdot10^{-5}$ & - \\ \hline
Gresho vortex density & -2 & \ref{eq:Lgresho} & $2.88\cdot10^{-2}$ & $2.96\cdot10^{-2}$ & $3.24\cdot10^{-2}$ & $4.11\cdot10^{-2}$\\ \hline
Gresho vortex pressure & -2 & \ref{eq:Lgresho}& $0.274$ & $0.277$ & $0.308$ & $0.352$\\ \hline
Gresho vortex velocity & -2 & \ref{eq:Lgresho}& $0.181$ & $0.169$ & $0.189$ & $0.176$\\ \hline
Noh density & -1/2 & \ref{eq:Lnoh}& $35$ & $41$ & $33.3$ & $38.8$\\ \hline
Noh pressure & -1/2 & \ref{eq:Lnoh}& $9.87$ & $10.1$ & $8.94$ & $9.86$\\ \hline
Noh velocity & -1/2 & \ref{eq:Lnoh}& $2.61$ & $2.29$ & $2.5$ & $2.24$\\ \hline
Simple waves density & -2 & \ref{eq:Lsimple}& $474$ & $135$ & $467$ & $133$\\ \hline
Simple waves pressure & -2 & \ref{eq:Lsimple}& $3.61\cdot10^{3}$ & $991$ & $2.75\cdot10^{3}$ & $914$\\
\hline
Simple waves velocity & -2 & \ref{eq:Lsimple}& $416$ & $99.4$ & $370$ & $97.9$\\ \hline
Shock tube density & -1/2 & \ref{eq:Lshock} &$7.35$ & $3.29$ & $7$ & $3.33$\\ \hline
Shock tube pressure & -1/2 & \ref{eq:Lshock} &$9.9$ & $9.56$ & $9.32$ & $9.26$\\ \hline
Shock tube velocity & -1/2 & \ref{eq:Lshock} &$3.74$ & $3.7$ & $3.52$ & $3.59$\\ \hline
Standing driven waves density & -2 & \ref{eq:Ldriven} &$6.3\cdot10^{-6}$ & - & $7.95\cdot10^{-6}$ & -\\ \hline
Standing driven waves pressure &  -2 & \ref{eq:Ldriven}& $1.06\cdot10^{-5}$ & - & $1.32\cdot10^{-5}$ & -\\
\hline
Standing driven waves velocity & -2 & \ref{eq:Ldriven}& $1.66\cdot10^{-5}$ & - & $8.02\cdot10^{-6}$ & -\\ \hline
\end{tabular}
\caption{The prefactors for the $L$ error function for different tests, as described in the text for our 1D and 2D test problems. The prefactor, $A$ is calculated from fitting the $L$ error function to the convergence rate $L=AN^\alpha$. Columns represent the reference to the equation in which $L$ is defined, the time advancement scheme and if the mesh points were allows to move in a semi-Lagrangian nature or not.}
\label{table:convergence}
\end{table}

\subsection{Kelvin Helmholtz Instability}
\begin{figure}
\begin{tabular}{c}
\includegraphics[width=0.4\textwidth]{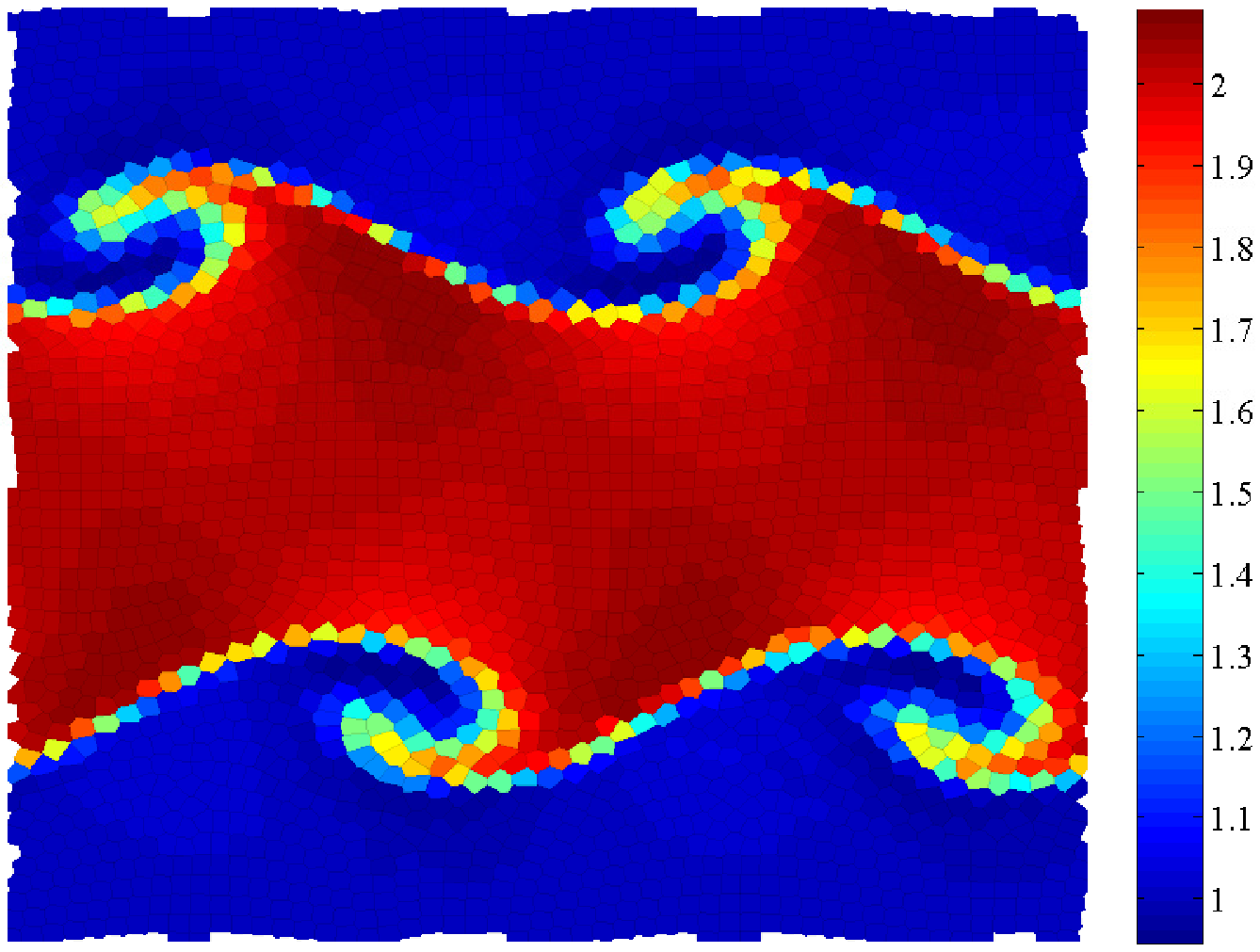} \\
\includegraphics[width=0.45\textwidth]{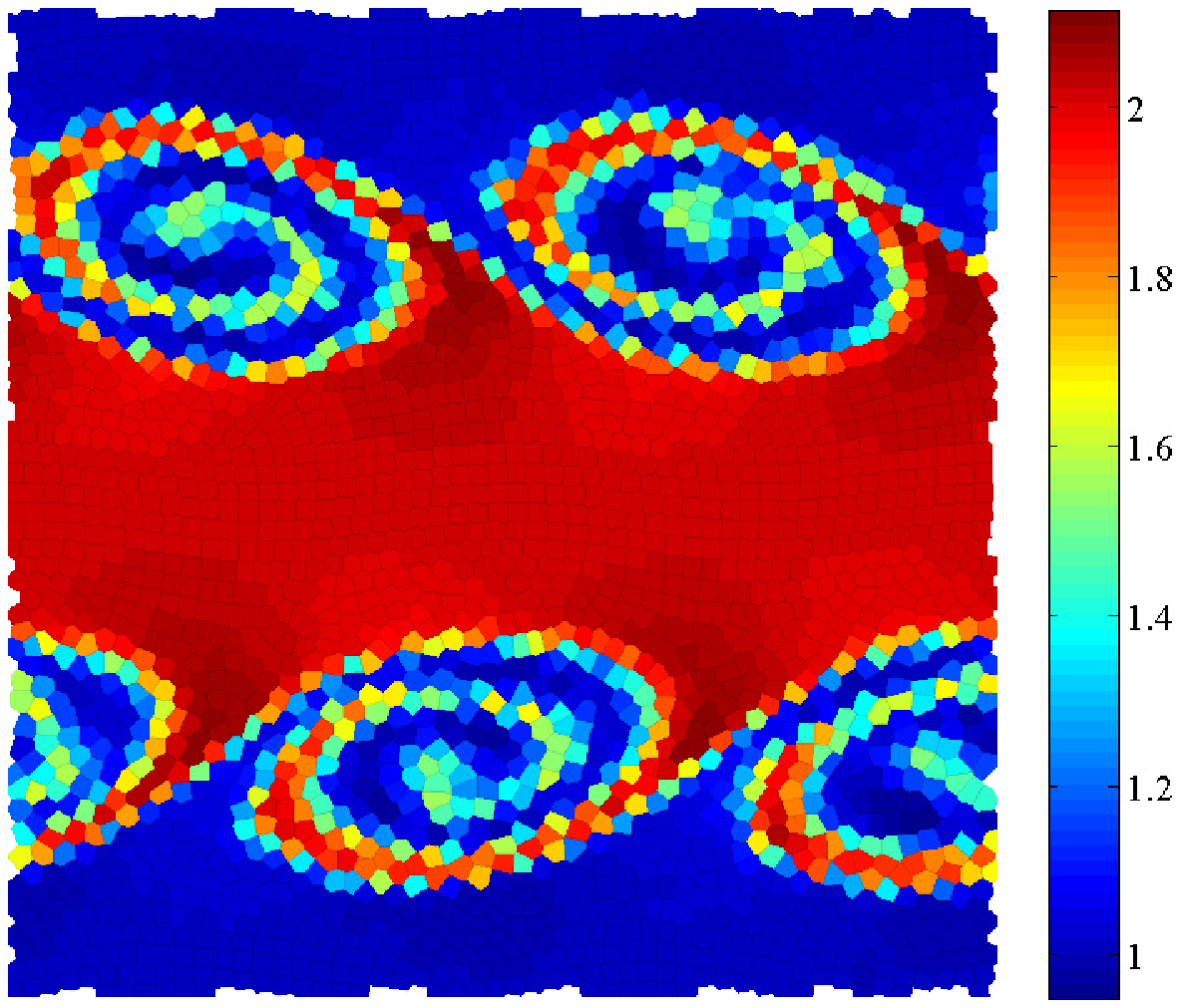}
\end{tabular}

\caption{Density map of the Kelvin Helmholtz problem at $t=1$ (left) $t=2$
  (right). The set up for this single mode test is described in the text.}

\label{fig:kelvin_helmholtz}

\end{figure}

One of the main benefits of a semi - Lagrangian
code is that it preserves contact discontinuities better than Eulerian
codes. A classic test that demonstrates this difference is the Kelvin
- Helmholtz instability \citep{Chandrasekhar1961}. This instability
occurs between two superposed fluid layers moving in parallel to their
interface. Our setup is identical to that described in AREPO, with
a resolution of $50\times50$ mesh generating points in the domain $[0,1]^2$ with periodic boundary conditions and a termination
time of $t=2$. Specifically the Pressure is set to be $P=2.5$ throughout the domain and the density and the velocity are given by
\begin{equation}
\rho\left(x,y\right)=\left\{ \begin{array}{c}
1\\
2
\end{array}\quad\begin{array}{c}
|y-0.5|>0.25\\
|y-0.5|<0.25,
\end{array}\right.
\end{equation}
\begin{equation}
v_x\left(x,y\right)=\left\{ \begin{array}{c}
-0.5\\
0.5
\end{array}\quad\begin{array}{c}
|y-0.5|>0.25\\
|y-0.5|<0.25,
\end{array}\right.
\end{equation}
\begin{equation}
v_y\left(x,y\right)=0.1\sin(4\pi x)\left( e^{-\frac{(y-0.25)^2}{2\sigma^2}}+e^{-\frac{(y-0.75)^2}{2\sigma^2}} \right)
\end{equation}
where $\sigma=0.05/\sqrt{2}$ and the adiabatic index is set to be $\gamma=5/3$.
Figure \ref{fig:kelvin_helmholtz} shows two snapshots
of our simulation at times $t=1$ and $t=2$. The snapshots seem very
similar to AREPO's (figure 32 in their paper) and preserve the discontinuity
between the fluids rather well. We also verified that these snapshots
do not change, even when a constant boost is added to all cells in
the initial conditions.

\subsection{Rayleigh Taylor Instability}

\begin{figure}[h]
\centering
\begin{tabular}{l c r}
  \includegraphics[width=0.1\textwidth]{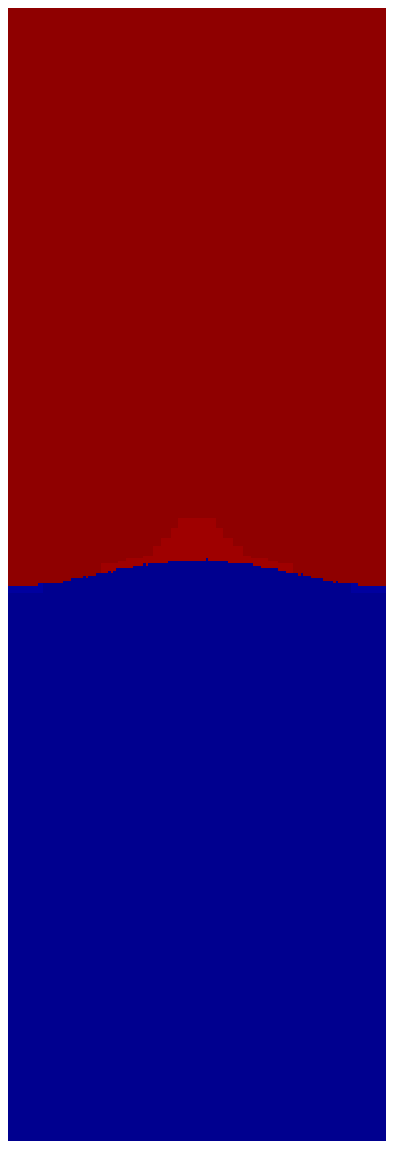} & 
  \includegraphics[width=0.101\textwidth]{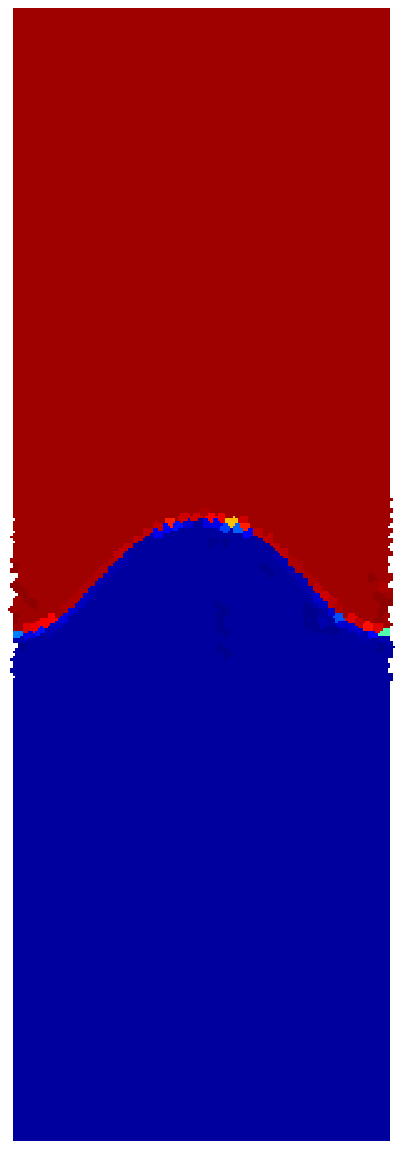} &
  \includegraphics[width=0.102\textwidth]{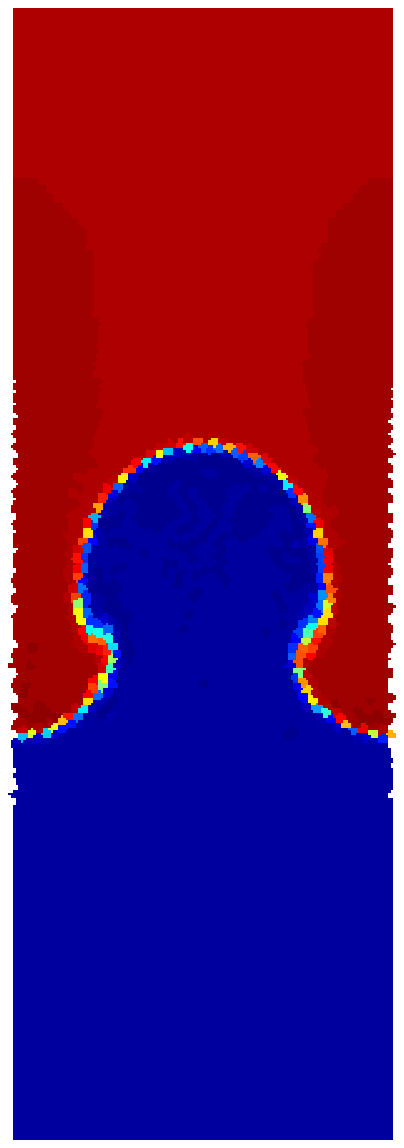} \\
  \includegraphics[width=0.1\textwidth]{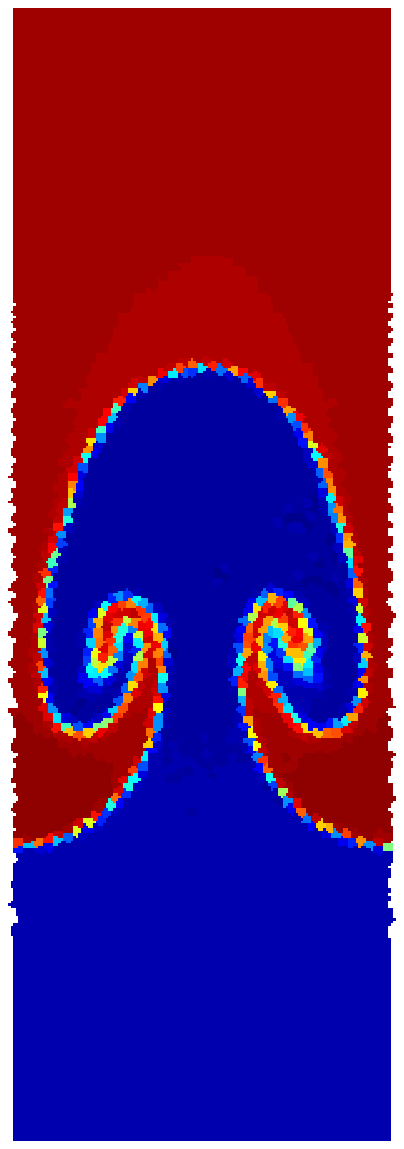} &
  \includegraphics[width=0.101\textwidth]{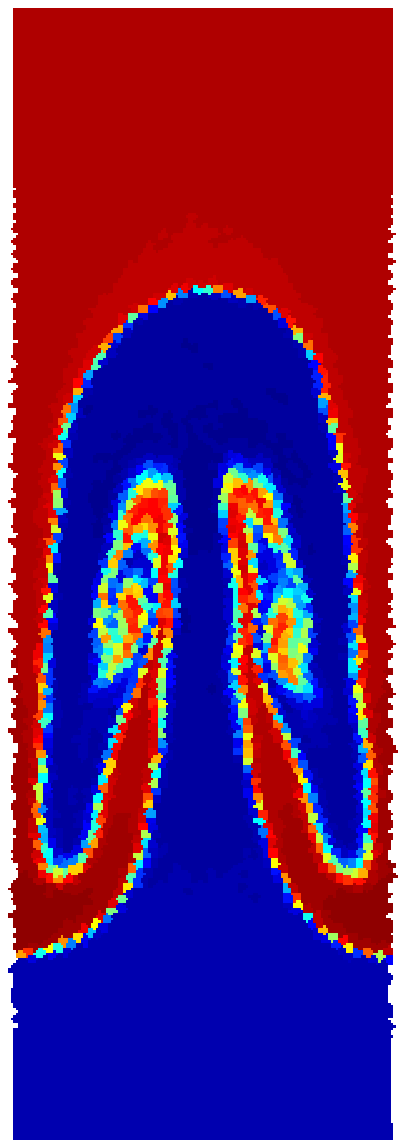} &
  \includegraphics[width=0.105\textwidth]{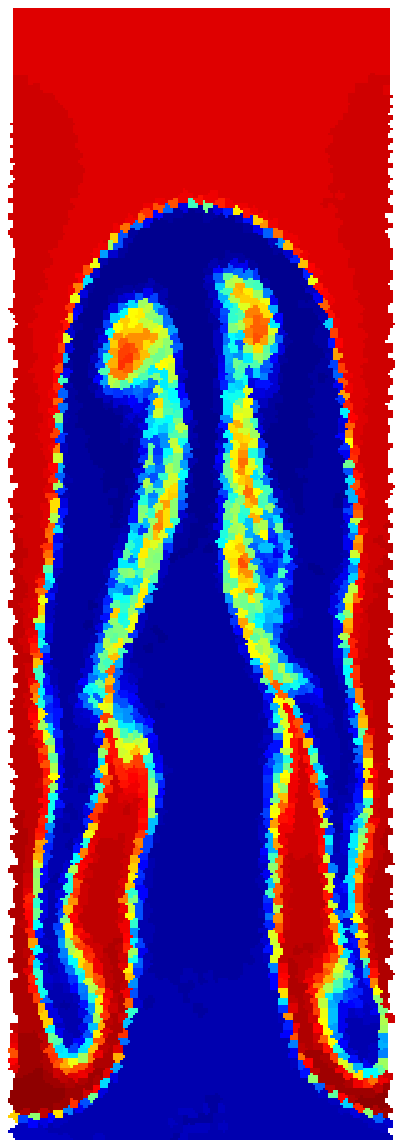}
\end{tabular}

\caption{Density maps of the Rayleigh Taylor instability at different times. The initial setup is given in the text.}

\label{fig:rayleightaylor}

\end{figure}

Rayleigh Taylor instability involves constant, uniform
external force and a discontinuity between densities. Again, the setup was the same as in AREPO \citep{Springel2010},
with a resolution $48\times144$ in the domain $x\in[0,0.5]$ and $y\in[0,1.5]$ and with periodic boundary conditions for the $x$ axis and reflecting for the $y$ axis. The initial setup is
\begin{equation}
v_x\left(x,y\right)=0,
\end{equation}
\begin{equation}
v_y\left(x,y\right)=w_0\left(1-\cos(4\pi x)\right)\left(1-\cos(4\pi y/3)\right),
\end{equation}
\begin{equation}
\rho\left(x,y\right)=\left\{ \begin{array}{c}
1\\
2
\end{array}\quad\begin{array}{c}
y<0.75\\
y>0.75,
\end{array}\right.
\end{equation}
\begin{equation}
P\left(x,y\right)=P_0+g\left(y-0.75\right)
\end{equation}
where $g=-0.1$, $P_0=2.5$ and $w_0=0.0025$. The simulation is run until $t=15$ with an adiabatic index $\gamma=1.4$.
run until time $t=15$. Figure
\ref{fig:rayleightaylor} shows snapshots at different times of this
simulation. As was shown in AREPO, semi-Lagrangian scheme is less diffusive than
the Eulerian scheme.

\subsection{Sod Shock Tube with Large Cell Volume Gradient}

There is a common knowledge that in AMR simulations, neighboring cells
should be of similar size \citep{Kravtsov1997}. In a moving mesh simulation,
cells of different sizes can become neighbors even without AMR. Neighbors
with large volume ratio can cause numerical errors in the simulation
and even crash it.

One reason for that, is that information travels
farther in large cells than in small cells. To demonstrate this phenomenon,
 the classic 1D Sod problem is run on a 2D grid with an uneven mesh.
The initial conditions are
\begin{equation}
\rho_{0}\left(x,y\right)=\left\{ \begin{array}{c}
1.0\\
0.125
\end{array}\quad\begin{array}{c}
y>0\\
y<0
\end{array}\right.
\end{equation}

\begin{equation}
p_{0}\left(x,y\right)=\left\{ \begin{array}{c}
1.0\\
0.1
\end{array}\quad\begin{array}{c}
y>0\\
y<0
\end{array}\right.
\end{equation}

\begin{equation}
v_{0}\left(x,y\right)=0
\end{equation}

\begin{figure}[h]
\centering
\begin{tabular}{l r}
\includegraphics[scale=0.5]{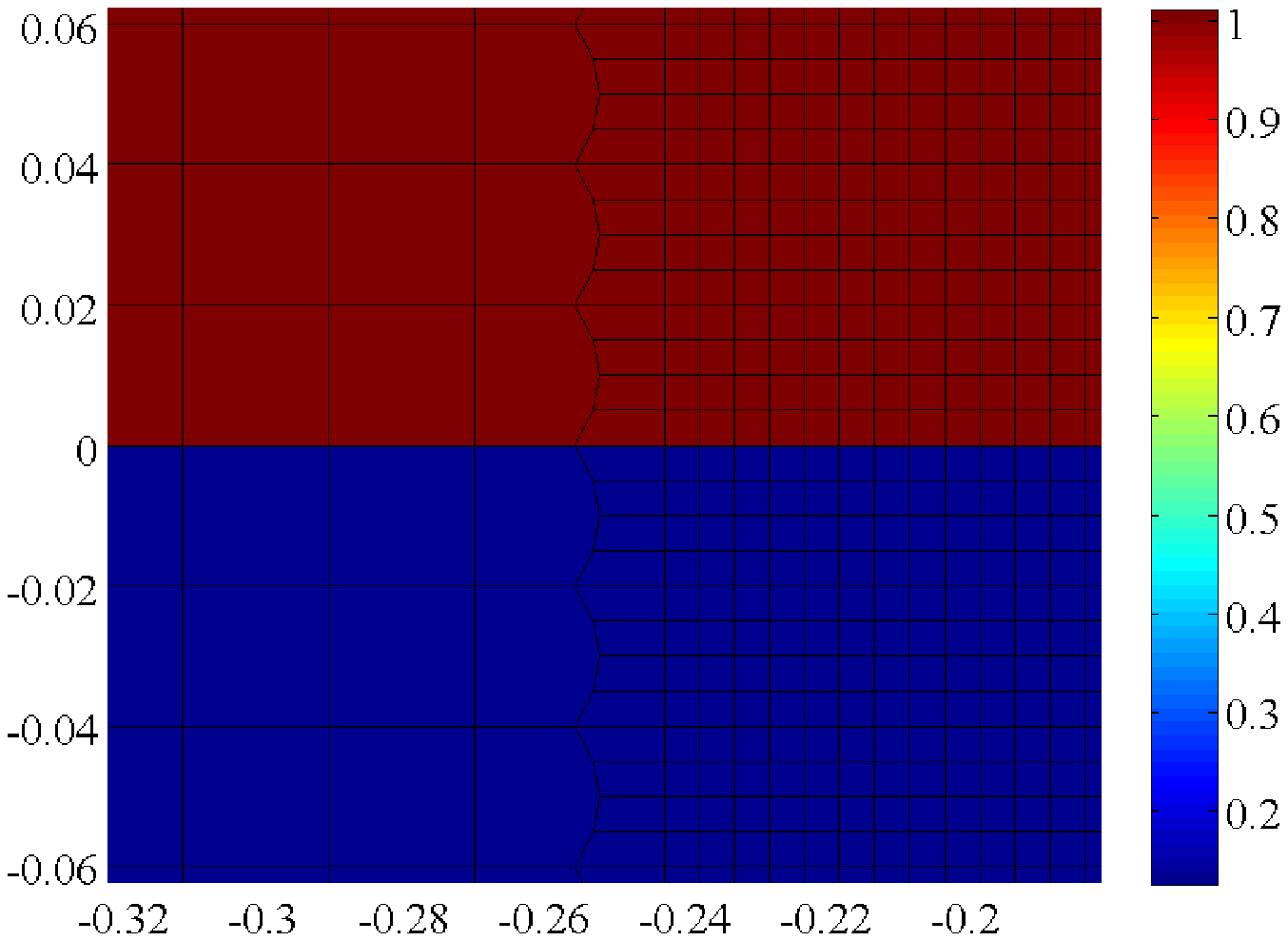} \\
\includegraphics[scale=0.5]{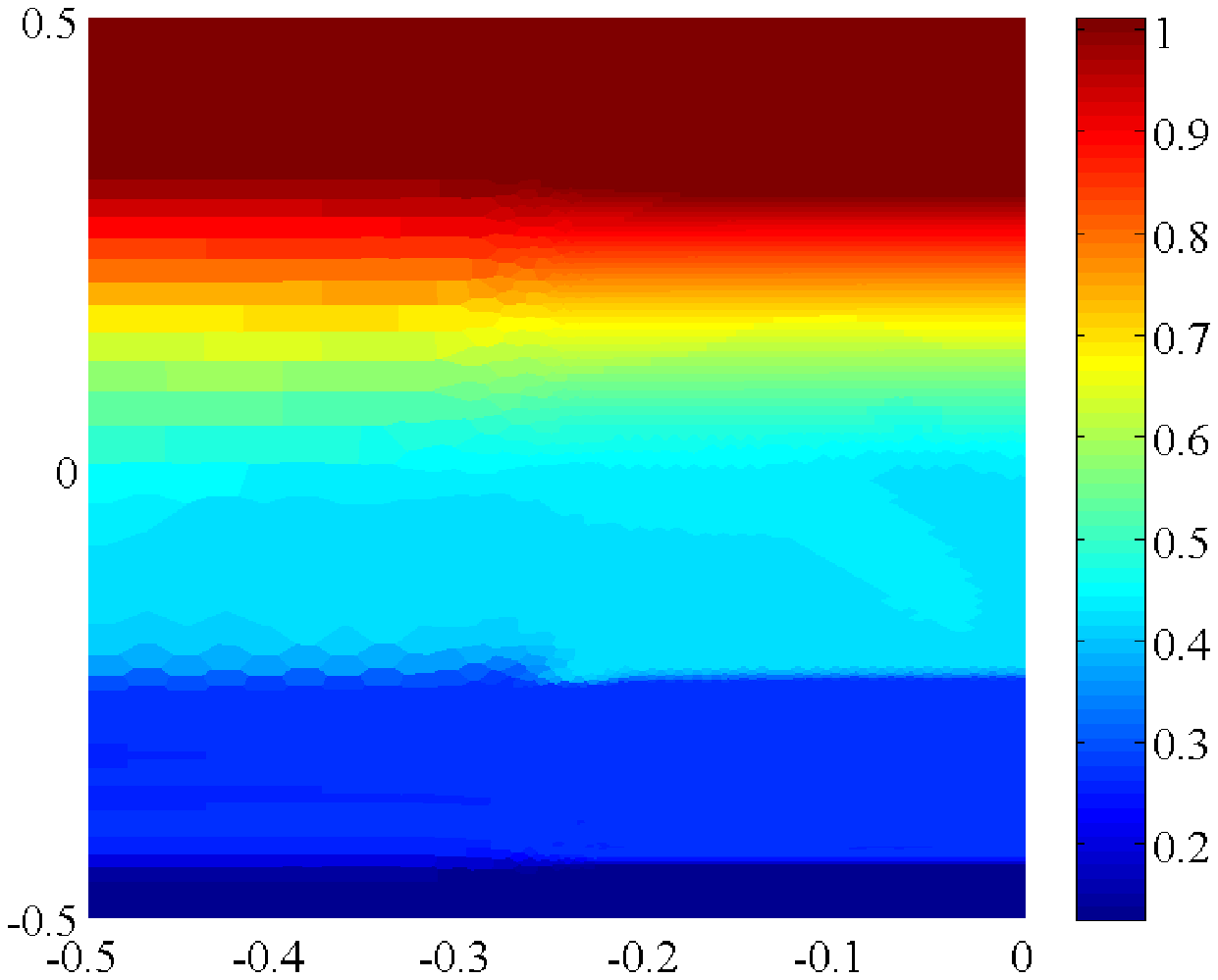}
\end{tabular}

\caption{The initial grid setup for the 2D Sod problem (left, zoomed in) and the evolved
symmetry breaking at $t=0.25$. Color denotes density. \label{fig:sod_2d}}
\end{figure}

  The initial conditions are independent of $x$, but the
resolution is not. The domain $0<x<-0.25$ has a resolution
of 25 cells, while the domain $-0.25<x<0$ has a resolution of
100 cells, as shown in figure \ref{fig:sod_2d}. As time advances,
the waves propagate at different velocities on each side of the grid
and that causes asymmetry in the hydrodynamics. Also,
small cells next to large cells tend to have aspect ratio much different
from unity, which can cause numerical errors and crash the simulation
in extreme cases.

  However, in most simulations this is not a critical issue
since the cell rounding scheme prevents most of the extreme cases of different size ratios. In the
few special cases where the cell rounding scheme does not fix this
issue, the problem can be remedied by splitting cells when they become
much larger than their neighbors, or coarsening cells when they get
much smaller than their neighbors.

\section{Area Inconsistency Problem}
\label{sec:error}
Since we solve the Riemann problem in a moving reference frame, this implicity assumes that the cell is going to change its area according to
\begin{equation}
\dot{A}_i=-\sum_{i\neq j}L_{ij}([\vec{w}_{j}-\vec{w}_{i}]\frac{\vec{c_{ij}}}{r_{ij}}-\frac{\vec{w}_{i}+\vec{w}_{j}}{2}\frac{\vec{r_{ij}}}{r_{ij}})
\label{eq:areachange}\end{equation}
where $\vec{w}$ is the velocity of the mesh generating point and the rest are defined in eq. \ref{eq:slope}.
However, the actual change in the cell's area is not $\dot{A}_i \Delta t$, which is accurate only to first order in time (more exactly to first order in the CFL number). The resulting difference between the expected change in the area and the actual change results in an error in the calculated fluxes (the scheme is still conservative). Moreover, the error is resolution independent since everything scales with the size of the cells and in principle can be of order unity.
This inconsistency can be demonstrated with a very simple test problem involving a strong shock, which induces a large variation in the cell's geometry. The initial conditions are set to be
\begin{eqnarray}
\rho&=&1\\
p&=&0.1\\
v_x&=&1\\
v_y&=&0
\end{eqnarray}
the adiabatic index is $\gamma=\frac{5}{3}$ and the problem is set up with a domain of $[-1,1]^2$
with rigid walls except the left wall which has inflow boundary conditions. The inflowing material creates a shock wave that moves to the left with a velocity of $U\sim0.448$ and has a post shock density of $\rho_d\sim3.23$. In the Lagrangian scheme, cells are compressed during their passage of the shock wave and the area inconsistency error is largest there. We run the simulation until $t=1.3$, with a CFL of 0.6, for various resolutions and record the maximal deviation from the analytical prediction in the downstream density (in units of $\rho_d$) as well as the error function that is defined as
\begin{equation}
L=\sum_{i=1}^N\frac{|\rho_i^n-\rho_d|}{\rho_d N}
\end{equation}
where the summation is done only on the downstream cells excluding those adjacent to the rigid wall and those adjacent to the shock wave.

\begin{figure}[h]
  \centering
  \includegraphics[width=0.4\textwidth]{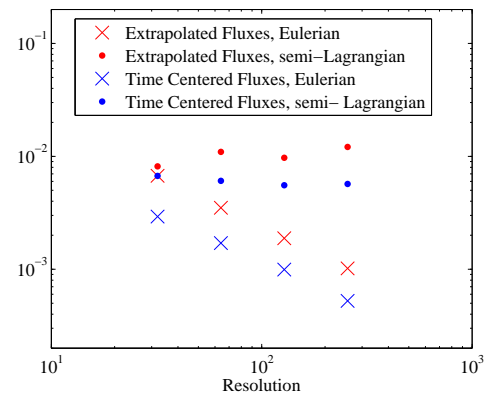}\\
  \caption{$L$, the error function of the density for the strong shock test, as a function of resolution for different time advancement schemes and for Eulerian or semi-Lagrangian meshes. The initial conditions for the test are described in the text.}\label{fig:l1res}
\end{figure}

\begin{figure}[h]
  \centering
  \includegraphics[width=0.4\textwidth]{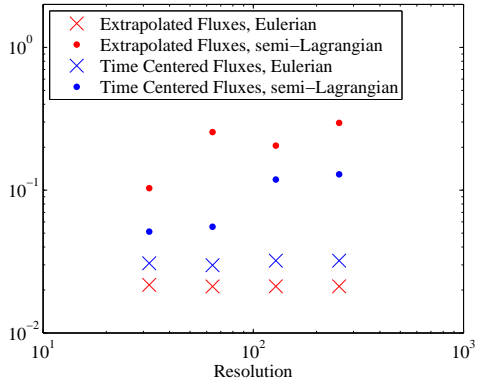}\\
  \caption{The maximal deviation from the analytical prediction in the downstream density (in units of $\rho_d$) as a function of resolution for different time advancement and for Eulerian or semi-Lagrangian meshes for the strong shock test. The initial conditions for the test are described in the text.}\label{fig:mdev}
\end{figure}

In figure \ref{fig:l1res} we show $L$, the error function, for the two time advancement schemes as well as for the Eulerian and Lagrangian point motion as a function of the resolution (the one dimensional number of points).
The nature of the time centered flux time advancement scheme solves eq. \ref{eq:areachange} to a higher accuracy than the extrapolated fluxes scheme.
The time centered flux time advancement scheme has an error that is a factor 2 less than than the extrapolated fluxes scheme. For lower CFL numbers, the ratio in the error between the two time advancement schemes only increases. Also, the error in the semi-Lagrangian schemes is constant due to the area inconsistency problem, while the Eulerian schemes have first order convergence, as expected. This is in stark contrast to the 1D Shock tube test, where the Lagrangian scheme was better since it had no area inconsistency problem.
The maximal error is indeed of order unity as can be seen in figure \ref{fig:mdev}. In fact the maximal error increases with resolution for the Lagrangian schemes, this is because there are more cells while the probability of having a large error is constant.

Since typically the errors between time steps are uncorrelated, the cumulative error is a random walk of the error of a single time step, until the error is large enough that it is canceled by the diffusion term.

Is this error critical? Typically large errors occur only when the cells are very ``unround", otherwise the difference between the calculated change in the area and the actual change are small. The errors do not change the overall dynamics of the simulation, but might cause errors on the level of a few percent in a few cells and in extreme cases error of order unity in a handful.

\section{Deviation from Lagrangian Motion and Diffusion}
\label{sec:nonlagrangian}
Moving the mesh generating points strictly with the fluid velocity can cause cells to become very elongated over time. This has the downside of causing the code to be unstable and even crash in extreme cases. An additional issue arises when two mesh points are close to each other, which can cause the mutual edge to have a large rotational velocity that can induce errors in the hydrodynamics. In order to fix this issue, AREPO has proposed to add an additional velocity to the mesh point whenever the mesh point is far from the center of the cell. The added velocity brings the mesh point closer to the cell's center. This fix, is controlled by two parameters, $\chi$, which defines in units of the cell's sound speed how fast the additional velocity is, and $\eta$, the criteria of how far in units of the cell's radius is the mesh point allowed to deviate from the cell's center before the fix is applied.

Since this additional velocity is typically not be in the direction of the fluid's velocity, a non-Lagrangian motion occurs, resulting with advection between cells. The more often the fix is applied, the more advection takes place, and the higher the fix velocity is, the higher the diffusion error in the advection term. However, having a diffusion error is not necessarily a bad thing since it allows to smooth out small scale errors. Mesh geometry induces errors with a wavelength comparable to the cell's size. The errors in the pressure and velocity tend to quickly adjust themselves to a smooth pattern while errors in the density take longer to smooth out if the motion is Lagrangian. An additional concern is that if $\chi\approx1$, it can have a negative effect on the time step since it can significantly increase the fluid's velocity relative to the edge's velocity.

\begin{figure}[h]
  \centering
  \includegraphics[width=0.4\textwidth]{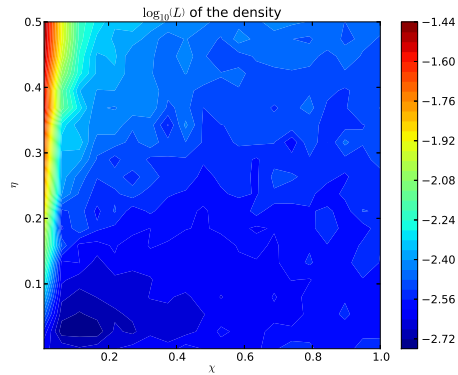}\\
  \caption{$L$, the error function of the density for the Gresho Vortex test for different values of $\chi$, the magnitude of the cell roundness fix in units of the cell's sound speed and $\eta$, the criteria how far the mesh point should be from the cell's center of mass in units of the cell's radius, in order to apply the fix. The initial conditions for the test and the definition of $L$ are described section \ref{sec:gresho}.}\label{fig:l_gresho}
\end{figure}

In order to show the dependence of the code on the fix parameters, we run the Gresho Vortex problem as presented in section \ref{sec:gresho} with different parameters of the cell roundness fix and with a resolution of $30^2$ cells. In figure \ref{fig:l_gresho} we show $L$, the error function as described in eq. \ref{eq:Lgresho} of the density for values of $\chi\in[0.01, 1]$ and $\eta\in[0.001, 0.5]$. The lowest value of $L$ is given approximately when $\chi=0.15$ and $\eta=0.02$, and we set those values to be our default choice when we run the code.

\section{Is Lagrangian Better?}
\label{sec:lagbetter}
In this section we focus on linear finite difference schemes, i.e.
recurrence relation of the form
\begin{equation}
y_{i}^{n+1}=\sum_{j}a_{j}y_{j}^{n}\label{eq:linear_scheme}
\end{equation}
where $y$ is the dependent variable, the lower index is the spatial,
the upper is temporal and $a_{j}$ are constants . Such schemes can
be solved analytically using Fourier transform \citep{Richtmyer1994}.
Since the hydrodynamic equations are non - linear, such scheme is
of little use. However, in the limit of small perturbations to a uniform background, it is
possible to obtain a linear approximation \citep{Landau1987}. In this
limit, the hydrodynamic equations are reduced to three decoupled linear
advection equations.
\begin{equation}
\frac{\partial s}{\partial t}+v_{0}\frac{\partial s}{\partial x}=0
\end{equation}
\begin{equation}
\frac{\partial j_{\pm}}{\partial t}+\left(v_{0}\pm c_{0}\right)\frac{\partial j_{\pm}}{\partial x}=0
\end{equation}
where $s=\log\left(p/\rho^{\gamma}\right)\approx\frac{\delta p}{p_{0}}-\gamma\frac{\delta\rho}{\rho_{0}}$
is the entropy and $j_{\pm}=\delta v\pm\frac{\delta p}{\rho_{0}c_{0}}$
are the Riemann invariants. In the limit of small perturbations Godunov's
method is reduced to a finite difference scheme with three decoupled
linear advection equations \citep{Toro1999}.

\subsection{First Order\label{sec:First-Order}}

The first order scheme for the linear advection equation
\begin{equation}
\frac{\partial y}{\partial t}+v \frac{\partial y}{\partial x} = 0
\end{equation}
(assuming positive drift velocity $v$) is
\begin{equation}
y_{i}^{n+1}=y_{i}^{n}+\xi\left(y_{i-1}-y_{i}\right)\label{eq:first_order_finite_difference_advection}
\end{equation}
where $\xi=\frac{v\Delta t}{\Delta x}$ is the Courant - Friedrichs
- Levy number, $\Delta x$ is the cell size and $\Delta t$ is the
time step. Denoting the imaginary number by $I=\sqrt{-1}$, to avoid confusion
with the indices, and $n=\frac{t}{\Delta t}=\frac{tv}{\Delta x\xi}$, we substitute the Fourier mode
\begin{equation}
y\left(x,t\right)=A\sigma^{t/t_{0}}\exp\left(-Ikx\right) \label{eq:fourier_mode}
\end{equation}
or equivalently
\begin{equation}
y_{i}^{n}=A\left(\sigma^{\Delta t/t_{0}}\right)^{n}\exp\left(Iki\Delta x\right)\label{eq:fourier_mode}
\end{equation}
into the first order finite difference scheme (equation \ref{eq:first_order_finite_difference_advection}) and have

\begin{equation}
\sigma^{\Delta t/t_{0}}=1-\xi\left(1-\exp\left(-Ik\Delta x\right)\right)
\end{equation}
\begin{equation}
\frac{y_{i}^{n}}{y_{i}^{0}}=\left[1-\xi\left(1-\exp\left(-Ik\Delta x\right)\right)\right]^{n}\label{eq:first_order_fft_mask}
\end{equation}
In the limit $\Delta x \ll \frac{1}{k}$ (while $\xi$ remains constant)
equation \ref{eq:first_order_fft_mask}) simplifies to
\begin{equation} \label{analytic_numeric_first_order_taylor}
\frac{y_{i}^{n}}{y_{i}^{0}}\approx\exp\left(-Iktv\right)\exp\left(-\frac{1}{2}\left(1-\xi\right)k^{2}\Delta xtv\right)\label{eq:first_order_fft_mask_approx}
\end{equation}
The first term on the right hand side is simply a shift (which happens
to be the exact Fourier filter for the advection equation), while
the second term is the leading term in the error caused by the finite
difference. In this case, the first order finite difference scheme
introduces artificial attenuation. We note that if $\xi>1$, then
attenuation becomes amplification, and the scheme becomes unstable.
If $\xi=1$, then the second term disappears and the wave travels
without distortions. This phenomenon is known as the ``magic time
step'' \citep{Taflove2005}. However, it is never used in practice,
mainly because, as we mentioned before, hydrodynamics involves three
wave speeds (which also vary in space), so it is impossible to choose a single time step for
which all CFL numbers would be 1.

\subsection{Second Order}
The same exercise as in section \ref{sec:First-Order} can be done
for a second order scheme, bearing in mind that it has to be second
order in both space and time.
\begin{equation}
y_{i}^{n+1/2}=y_{i}^{n}+\frac{1}{4}\xi\left(y_{i-1}^{n}-y_{i+1}^{n}\right)
\end{equation}
\begin{equation}
y_{i}^{n+1}=y_{i}^{n}+\frac{1}{2}\xi\left(y_{i-1}^{n+1/2}-y_{i+1}^{n+1/2}\right)
\end{equation}

Substituting the Fourier mode (equation \ref{eq:fourier_mode})
yields the filter
\begin{equation} \label{eq:second_order_filter_unabridged}
\frac{y_{i}^{n}}{y_{i}^{0}}=\left(1-I\xi\sin\left(k\Delta x\right)-\frac{1}{2}\xi^{2}\sin^{2}\left(k\Delta x\right)\right)^{n}
\end{equation}
In the limit of small $\Delta x$
\begin{equation} \label{eq:second_order_scheme_sigma_small_dx}
\frac{y_{i}^{n}}{y_{i}^{0}}\approx\exp\left(-Ikvt\right)\exp\left(\frac{I}{6}\left(1-\xi^{2}\right)k^{3}\Delta x^{2}tv\right)
\end{equation}
In this case, the leading error decreases the effective propagation
speed, so the numerical wave always lags behind the exact solution.
This scheme is unstable for all values of the CFL number, in accordance
with Godunov's theorem \citep{Godunov1961}. In practice, this difficulty
is circumvented by the use of slope limiters \citep{Toro1999}, which
introduce non linearity to the scheme.

\subsection{Grid Motion}
\begin{figure}
\centering
\includegraphics[width=0.4\textwidth]{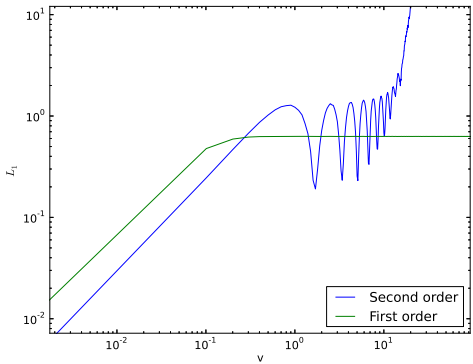}
\caption{The variation of the $L_1$ norm (see equation \ref{eqn:l1_transport_equation})
relative to the analytic solution as a function of the drift velocity $v$, for
both first and second order schemes of the linear advection equation. The resolution is 100 cells, the initial profile was a sine wave with amplitude 1 and wavenumber $2 \pi \cdot 10$, the time is 1 and the CFL coefficient is 0.3 .\label{fig: l1_vs_v}}
\end{figure}

The formalism described above can be used to explore
the effects of grid motion (which is usually chosen to be either Eulerian
or Lagrangian) by repeating the calculation described above but varying the ambient velocity. In a simple linear advection equation, the higher the velocity
the less accurate the scheme will be. We recall that a snapshot of some variable is represented by a discrete set of values at fixed position. Suppose we start out with the same initial condition, and advance it to time $t$ using two different methods. The first method is using the exact solution to the advection equation, and the second method is using the analytic - numeric method describe above. This will yield two sets of values $\phi^1_i$ and $\phi^2_i$, where $i$ is the spatial index. In order to measure how close both sets are, we define the following function
\begin{equation} \label{l1_norm_def}
 L_{1} = \frac{1}{AN}\cdot\sum\limits_{i=1}^{N}\left|\phi^1_{i}-\phi^2_{i}\right| \label{eqn:l1_transport_equation}
\end{equation}
where $N$ is the number of terms of $x_i$ (and also $y_i$) and $A$ is the amplitude of the wave. The latter is included so that $L_1$ will be dimensionless.
Figure \ref{fig: l1_vs_v} shows the variation of
the $L_1$ error norm as a function of the ambient velocity for both first and second order time advance schemes for the case of a single mode as the initial condition, where the resolution is 100 cells, the wavenumber is
$2 \pi \cdot 10$, the time is 1 and the CFL number is 0.3.
The first order scheme seems to grow linearly, and then saturates. This occurs when the wave decays to zero due to numerical viscosity, so $L_1 \le 1$.
In the case of the second order scheme, the error first increases, but then starts
decreasing and continues to oscillate. The oscillations occur since the lag increases with
the velocity, but when the phase approaches a multiple of $2 \pi$ the numeric and
analytic waves coincide and the error decreases. At even higher velocities the next
term dominates and the error grows monotonically.
$L_{1}$ at early times for both schemes can be approximated analytically. We assume that the initial conditions are a single Fourier mode $y(x) = \exp \left( I k x \right)$. We then obtain two spatial profiles at a later time time $t$. The first profile is obtained using the exact solution to the advection equation by multiplying by $\exp \left(-I k t v \right)$. The second profile is obtained by multiplying by the filter of the first order scheme, taylor expanded for $\Delta x \rightarrow 0$ (equation ~\ref{analytic_numeric_first_order_taylor}).
Comparing the two profiles using the $L_1$ norm (equation ~\ref{l1_norm_def}) yields
\begin{center}
 $L_{1}^{1o} = \left(1 - e^{ -\frac{1}{2} \left( 1 - \xi \right)k^2 \Delta x t v} \right) \frac{k}{\pi} \int^{\pi/k}_{0} \sin \left( k x \right) dx=$
\end{center}

\begin{equation}
 = \frac{2}{\pi} \left[1 - \exp \left( -\frac{1}{2} \left( 1 - \xi\right) k^2 \Delta x t v \right) \right]
\end{equation}

We assumed that the amplitude was positive, so that the analytic solution would always be
larger than the numeric, and thus the absolute value can be dropped. In principle, the integration
should be carried out in the range $[0,2\pi/k]$ (i.e. over one cycle) but due to the symmetry,
suffice to integrate over the range $[0,\pi/k]$.
A similar calculation can be performed for the second order scheme. Again, we start out with a pure Fourier mode as initial conditions $y(x) = \exp \left(I k x \right)$. We obtain two profiles at a later time $t$: once using the exact filter $\exp \left(-I k t v \right)$ and a second time using the filter for the second order scheme, Taylor expanded about $\Delta x \rightarrow 0$ (equation ~\ref{eq:second_order_scheme_sigma_small_dx}). In this case, the numeric solution lags behind the analytic solution.
\begin{center}
 $L_{1}^{2o} = \frac{k}{2\pi} \int_0^{2\pi/k} \left| \sin \left(k \left(x-v t \right) \right) - \sin \left(k \left( x - v_n t\right) \right) \right| dx = $
\end{center}

\begin{equation}
 = \frac{4}{\pi} \left| \sin \left( k t \left( v - v_n \right) \right) \right|
\end{equation}

Where $v$ denotes the drift velocity and $v_n=v \left( \frac{1}{6} \left(1-\xi^2 \right) k^2 \Delta x^2 \right) $
is the numeric velocity.
In contrast to the monotonous behavior of the first order scheme, the $L_1$ of the second order oscillates, since the lag between the waves increases until the phase difference is $2\pi$. At that point the error drops to zero, and the cycle repeats itself. The reason for this counter intuitive behavior is that this based on tailor expansion for small $t$ (equation ~\ref{eq:second_order_scheme_sigma_small_dx}). At larger values of $t$ this approximation no longer holds, and one must resort to the complete expression for the second order filter (equation ~\ref{eq:second_order_filter_unabridged}).

\begin{figure}
\centering
\includegraphics[width=0.4\textwidth]{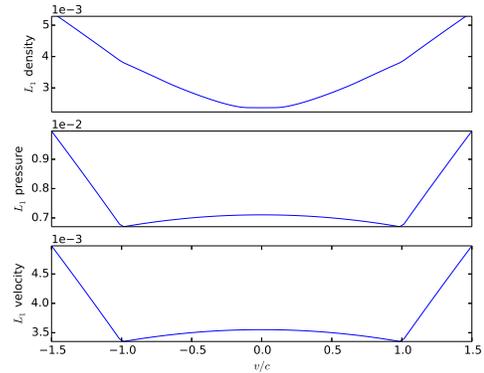}
\caption{$L_1$ measure of the difference between analytic and numerical solutions as a function of the Mach number for the first order scheme. The initial conditions for this problem were: $\rho(x) = 1$, $p(x) = 1 + 10^{-3} \exp\left(-\frac{(x-1/2)^2}{0.001}\right)$ and $v(x) = 0$. The number of points is 1000, the CFL number was 0.3, and the simulation was carried on to time $t=0.1$ \label{fig: l1_vs_v2c_first_order}}
\end{figure}

In order to demonstrate the effect of drift velocity on the accuracy of a finite difference scheme, we performed the following test. We used the same initial conditions for the perturbations ($\delta \rho = 0$, $\delta p = 10^{-3}\exp\left(-\frac{(x-1/2)^2}{0.001}\right)$, $\delta v = 0$) and changed the drift velocity. For every velocity we advanced the hydrodynamic profiles to a time $t = 0.1$ using the analytic formalism described above, and compared the result to the analytic profiles using the $L_1$ metric. The results for a first
order scheme are presented in figure \ref{fig: l1_vs_v2c_first_order}.
The minima occur whenever one of the wave speeds becomes zero. Since
the velocity and pressure propagate only through sound waves, their
minima occur at $v=\pm c$. In the case of density, the dominant contribution
is from the entropy wave, a minimum only occurs at $v=0$.

\begin{figure}
\centering
\includegraphics[width=0.4\textwidth]{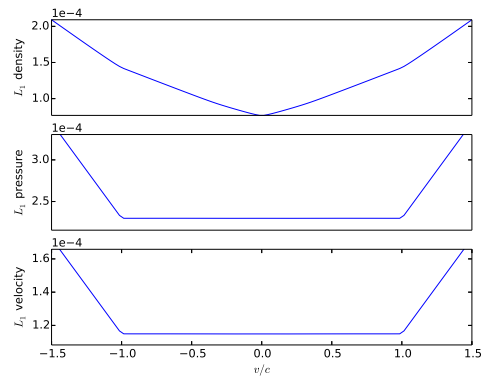}
\caption{$L_1$ measure of the difference between analytic and numerical solutions as a 
  function of the Mach number for the second order scheme. 
  The problem considered here is the same as in figure ~\ref{fig: l1_vs_v2c_first_order} \label{fig: l1_vs_v2c_second_order}}
\end{figure}

  The same behavior recurs in second order schemes, as can
be seen in figure \ref{fig: l1_vs_v2c_second_order}. The reason for
the plateau in the range $v\in\left[-c,c\right]$ is that the errors
from the left and right sound waves exactly balance each other.

These results show us that in general, a Lagrangian grid
will not always give better results than an Eulerian grid.

\section{Conclusion}{\label{sec:Conclusion}

We presented our version of a hydrodynamic code on a moving Voronoi
mesh. This code is similar to AREPO, with several important exceptions of a
few implementation details. Our code, in its current initial form, still lacks some features
available in AREPO. These features include three dimensional geometry
and individual time steps.

  With our new code we explored the question whether a simulation
based on a moving mesh gives better results than static mesh. In our
array of tests, Lagrangian grid tends to give better results than
an Eulerian grid. However, a more detailed one dimensional analysis
reveals some scenario where an Eulerian grid would surpass a Lagrangian grid.

Comparing the different time advancement schemes between the codes show that for purely hydrodynamic problems, AREPO's scheme tends to give slightly better results for small perturbations, while for external sources our time advancement scheme gives better results for pressure and density, and AREPO does better for the velocity.

\noindent
Our code is publicly available at \\
\href{https://code.google.com/p/huji-rich/}{https://code.google.com/p/huji-rich/}. 
The open source nature of our code allows other users to both reproduce
the results presented here and run the code for their own calculations.
Our code is built in a modular object oriented fashion to allow other users to incorporate new physics with
ease.
\begin{acknowledgments}
We would especially like to thank Omer Bromberg and Orly Gnat who wrote the HLLC solver and gave valuable advice. We would also like to thank Udi Nakar and Tsvi Piran for their insightful comments regarding writing the code. ES is supported by an Ilan Ramon grant from the
Israeli Ministry of Science. This research is supported in part by  ISF, ISA, iCORE grants and a Packard Fellowship.
\end{acknowledgments}

\bibliographystyle{apj}
\bibliography{rich_paper}

\end{document}